\documentstyle[12pt]{article}
\def\beq{\begin{equation}}
\def\eeq{\end{equation}}
\def\bea{\begin{eqnarray}}
\def\eea{\end{eqnarray}}
\def\bq{\begin{quote}}
\def\eq{\end{quote}}

\def\PLB{{\it Phys. Lett.} }

\def\PR{{\it Phys.Rev.} }

\def\gappeq{\mathrel{\rlap {\raise.5ex\hbox{$>$}}
{\lower.5ex\hbox{$\sim$}}}}

\def\lappeq{\mathrel{\rlap{\raise.5ex\hbox{$<$}}
{\lower.5ex\hbox{$\sim$}}}}

\begin{document}
\topmargin -0.5cm
\oddsidemargin -0.3cm
\pagestyle{empty}
\begin{flushright}
{CERN-TH.7518/94}
\end{flushright}
\vspace*{5mm}
\begin{center}
{\bf THE RENORMALIZATION  GROUP INSPIRED APPROACHES }\\
{\bf AND ESTIMATES OF THE TENTH-ORDER  CORRECTIONS} \\
{\bf TO THE MUON ANOMALY IN QED }\\
\vspace*{1cm}
{\bf Andrei L. Kataev} \footnote{On leave of absence from
Institute for Nuclear Research of the Academy of Sciences
of Russia, 117312 Moscow, Russia; address after 1 January
1995}
\\
\vspace{0.3cm}
Theoretical Physics Division, CERN, \\
CH-1211 Geneva 23, Switzerland \\
\vspace{0.5cm}
and \\
\vspace*{0.5cm}
{\bf Valery V. Starshenko} \footnote{Based on the research done
for a PhD thesis at the Moscow State University}.
\\
\vspace*{0.3cm}
%Chair of General Physics, \\
%Zaporozhye Industrial Institute \\
Aschara, Germany \\
\vspace*{1cm}
{\bf ABSTRACT} \\
\end{center}
\vspace*{2mm}
\noindent
We present the estimates of the five-loop QED corrections
to the muon anomaly using the scheme-invariant approaches
and demonstrate that they are in good agreement with the
results of exact calculations of the corresponding tenth-order
diagrams supplemented by the additional guess about the
values of the non-calculated contributions.
\vspace*{3cm}
\begin{flushleft}
CERN-TH.7518/94\\
December 1994
\end{flushleft}
\newpage
\setcounter{page}{1}
\pagestyle{plain}
%\section{Introduction}
{\bf I. Introduction}

 The direct
analytical or numerical calculations of the higher-order terms
to the physical quantities
 in the concrete
renormalization schemes provide the important information
about the behaviour of the corresponding perturbative
approximations.
 However, there are also some other
approaches to treat the problem of the extraction of
certain information from the truncated perturbative series.
These approaches are the principle of minimal sensitivity (PMS)
\cite{PMS} and the effective charges (ECH) prescription \cite{ECH},
which is equivalent {\it a posteriori} to the scheme-invariant
perturbation theory \cite{SIP}. Of course, it is better to use
these approaches directly in the concrete orders of the perturbation
theory, as was done in QCD in Refs. \cite{GKLS}-\cite{ChK}.
 However, if one adopts the point of view
that these methods really pretend to the role of  ``optimal''
procedures in the sense that they might provide better convergence
of the corresponding approximations in the non-asymptotic regime,
it is possible to try to go one step further and apply the
procedure of re-expansion of the ``optimized'' expressions in
the coupling constant of an initial scheme. One can consider
 the residual ($N$+1)-th order term as the estimate of the
($N$+1)-th order correction in the initial scheme \cite{PMS}.

The re-expansion procedure was already applied for the analysis
of the perturbative predictions for  $(g-2)_{\mu}$ in
QED \cite{PMS,KS} (for  related considerations see Ref. \cite{F})
and for the estimates of the QCD corrections to definite physical
quantities. In these works, the
quantities under study are the Drell-Yan cross-section
at the $O(\alpha_s^2)$-level \cite{DY}, $R(s)=\sigma_{tot}(e^+e^-
\rightarrow hadrons)/\sigma(e^+e^-\rightarrow\mu^+\mu^-)$,
$R_{\tau}=\Gamma(\tau\rightarrow \nu_{\tau}+hadrons)/\Gamma(\tau
\rightarrow\nu_{\tau}\overline{\nu}_ee)$, non-polarized and polarized
Bjorken sum rules at the $O(\alpha_s^4)$ and even $O(\alpha_s^5)$-
levels \cite{KatSt1,KatSt2} and the singlet contribution to the
Ellis-Jaffe sum rule at the $O(\alpha_s^3)$-order \cite{Kat2}.

It is clear that the re-expansion formalism, which is similar
to the procedure used in Ref. \cite{BarRem}
to predict the RG-controlable
$ln(m_{\mu}/m_e)$-terms from the expression for $(g-2)_{\mu}$
through the effective coupling constant
$\overline{\alpha}(m_{\mu}/m_e)$,  correctly reproduces
the RG-controlable terms \cite{PMS}, \cite{FS}.
 One can also hope that it can give
the impression about the possible values of the constant terms
as well. This hope is based on the observation of the
existence of a satisfactory agreement of the results of application
of the re-expansion procedure in QED \cite{KS} and QCD
\cite{KatSt1,KatSt2}
with the results of the explicit calculations.
It should be stressed that on the contrary to the RG
considerations
of Ref. \cite{BarRem}, the ``optimization methods''
are dealing with the full RG-invariance of the quantities
under consideration, which produce the additional equations,
relevant to the freedom in the choice of the higher-order
coefficients of the $\beta$-function. The solution of these
equations gives the possibility to define the sets of scheme-invariants
\cite{PMS} which are the cornerstones of the ``optimization''
methods.

However, in the
definite cases the procedure of re-expansion of the ``optimized''
results
can run against some reef, which was
overlooked in the process of some previous applications
\cite{PMS,KS,F}. In the case of the analysis of the perturbative
series for $(g-2)_{\mu}$ this problem grows out from the
non-careful treatment of the light-by-light scattering graphs
with the electron loop coupled to the external photon line.

In Sec.I of this work we repeat the decription of the
basis of the formalism used by us.
 The exact expressions
for the terms in the re-expansion formulas are derived.
It is demonstrated that the estimates obtained using the
re-expansion of the ECH expressions are identical to the
results of calculations of the ($N$+1)-th order corrections in the
special scheme, where all lower order coefficients of the
physical quantities and the QCD $\beta$-function are defined in
a certain fixed scheme (in the case of QED the on-shell (OS)
scheme is usually used) and the ($N$+1)-th order coefficient of
the $\beta$-function coincides with the ($N$+1)-th order scheme-invariant
coefficient of the ECH $\beta$-function $\beta_{eff}$.

In Sec.II, using the information about the four-loop
coefficient of the QED $\beta$-function in the OS scheme
\cite{BrKT} we generalize the considerations of Refs. \cite{PMS,KS,F}
to the five-loop level.
 We  follow the proposals
of Ref. \cite{BLM} and  consider the light-by-light scattering
 graphs mentioned above separately
in our  RG-inspired analysis. We show that this empirical
improvement leads to more satisfactory and thus more reliable
estimates of the five-loop contributions to $(g-2)_{\mu}$ than
in the case of non-separation of the light-by-light scattering
contributions. Finally in the Appendix we present the expressions
for  the six-loop RG-controllable contributions to the muon
anomalous magnetic moment which follow from the analysis of Sec.II.

{\bf II. The Description of the Formalism}

Consider first the order $O(a^{N})$ approximation of a Euclidean
renormalization group invariant quantity
\beq
D_{N} = d_{0} a(1 + \sum^{N-1}_{i=1} \, d_{i} a^{i})
\label{1}
\eeq
with $a = \alpha_{s}/\pi$ being the solution of the corresponding
renormalization group equation for the $\beta$-function which is
defined as
\beq
\mu^{2} \frac{\partial a}{\partial \mu^{2}} =
\beta (a) = - \beta_{0} a^{2}
(1 +  \sum^{N-1}_{i=1} \, c_{i} a^{i})\ .
\label{2}
\eeq
 The coefficients
$d_{i}, i \geq 1$ and $c_{i}, i \geq 2$ are scheme-dependent.
In order to calculate them in practice it is necessary to specify
the scheme of subtractions of the ultraviolet divergences.
In QED the OS scheme is commonly used.
 However, this scheme is not the unique prescription for
fixing the RS ambiguities, which affect the values of these coefficients.
In both phenomenological and theoretical
studies other methods are also widely applied.

The PMS \cite{PMS} and ECH \cite{ECH} prescriptions stand out from
various methods of treating scheme-dependence ambiguities.
 Indeed, they are based on the conceptions of the
scheme-invariant quantities, which are defined as the combinations of
the scheme-dependent coefficients in Eqs. (1) and (2).  Both these
methods pretend to the role of
  ``optimal" prescriptions, in the sense that they might
 provide better convergence of the corresponding approximations in
the non-asymptotic regime, and thus allow an estimation of the
uncertainties of the perturbative series
in the definite order of perturbation theory.  Therefore,
applying these
 ``optimal" methods  one can try to estimate the
effects of the order $O(a^{N+1})$-corrections starting from the
approximations $D^{opt}_{N} (a_{opt})$ calculated in a certain
``optimal" approach \cite{PMS}, \cite{KS}, \cite{FS}.

Let us follow the considerations of Ref. \cite{PMS}
re-expand $D_{N}^{opt} (a_{opt})$ in terms of the coupling
constant $a$ of the particular scheme
\beq
D_{N}^{opt} (a_{opt}) = D_{N} (a) + \delta D_{N}^{opt} a^{N+1}
\label{3}
\eeq
where
\beq
\delta D_{N}^{opt} = \Omega_{N}(d_{i}, c_{i}) -
\Omega_{N} (d_{i}^{opt}, c_{i}^{opt})
\label{4}
\eeq
are the  numbers which simulate the coefficients of the
order $O(a^{N+1})$-corrections to the physical quantity, calculated in
the particular initial scheme.  The
coefficients $\Omega_{N}$ can be obtained from the following system of
equations:
\bea
\frac{\partial}{\partial \tau}
(D_{N} + \Omega_{N} a^{N+1}) =
O(a^{N+2}), \nonumber \\
\frac{\partial}{\partial c_{i}}
(D_{N} + \Omega_{N} a^{N+1}) =
O(a^{N+2}),\ i \geq 2
\label{5}
\eea
where the parameter $\tau = \beta_{0} \ell n (\mu^2 / \Lambda^2)$
represents  freedom in the choice of the renormalization point
$\mu$. The conventional scale parameter $\Lambda$ will not explicitly
appear in all our final formulas. The system of these
equations can be solved following the lines of ref. \cite{PMS}.
Let us stress again that the difference between the ``optimization''
equations
and the RG-approach of Ref. \cite{BarRem} lies in the fact that
the latter one is dealing with the first equation from the
the system of Eq.(\ref{5}) only.
The quantities $\Omega_{l}$
%and $\Omega_{3}$ in which we will be interested
can be related to the scheme invariants $\rho_{l}$ in the
following way:
\beq
\rho_{l}=d_l+\frac{1}{l-1}c_l-\Omega_{l}(d_1,...,d_{l-1}
;c_1,...,c_{l-1}) .
\label{Omega}
\eeq
Note that the general expressions of the scheme-invariants $\rho_{l}$
and of the correction terms
$\Omega_l$ can be defined in  different ways.
 Various definitions differ by  scheme-independent constant terms.
We are choosing these correlated constant terms imposing the condition
 that the expressions
for the scheme-invariants $\rho_l$ are connected with the coefficients
 $c_l^{ECH}$ of the ECH $\beta$-function
\beq
\beta_{eff}(a_{ECH})=-\beta_0a_{ECH}^2\bigg(1+c_1a_{ECH}+
\sum_{i\geq 2} \, c_{i}^{ECH}a_{ECH}^{i}\bigg )
\label{ebeta}
\eeq
as
\beq
\rho_l= \frac{c_{l}^{ECH}}{l-1}
\label{defrho}
\eeq
where
\beq
D(a_{ECH})=d_0a_{ECH}(a) \ .
\label{ech}
\eeq
The concrete expressions for
the invariants $\rho_l$ and thus for
the correction terms     $\Omega_l$
can be derived from the following equation:
\beq
\beta_{eff}(a_{ECH})=\frac{\partial a_{ECH}}{\partial a}\beta(a) .
\label{rel}
\eeq

 We
present here  the final expressions, which are already known \cite{PMS}:
\beq
\Omega_{2} = d_{0}d_{1} (c_{1} + d_{1}),
\label{6}
\eeq
\beq
\Omega_{3} = d_{0}d_{1} (c_{2} - \frac{1}{2} c_{1}d_{1}
-2d_{1}^{2} + 3d_{2})\ .
\label{7}
\eeq
and the new  terms which we evaluated
\beq
\Omega_{4}=\frac{d_0}{3} ( 3c_{3}d_1+c_2d_2-4c_2d_1^2+2c_1d_1d_2
-c_1d_3+14d_1^4-28d_1^2d_2+5d_2^2+12d_1d_3 )
\label{new1}
\eeq
\begin{eqnarray}
\Omega_5&=&\frac{d_0}{4} (4c_4d_1-8c_3d_1^2+2c_3d_2-4c_2d_1d_2
+8c_2d_1^3-2c_1d_4+6c_1d_1^4
-16c_1d_1^2d_2+3c_1d_2^2 \nonumber \\
&&+8c_1d_1d_3
-48d_1^5+120d_1^3d_2-48d_1d_2^2+16d_2d_3-56d_1^2d_3+20d_1d_4).
\label{new2}
\end{eqnarray}
These terms reproduce
 the RG controllable logarithmic contributions. In the case
of the five-loop level one can reobtain the QED results
presented in Ref. \cite{Kat3}. We discuss this point in more
detail in the next Section.

It should be stressed that in the ECH approach $d_{i}^{ECH} \equiv 0$
for all $i \geq 1$.  Therefore one gets the following expressions for
the higher-order corrections in Eq. (3):
\beq
\delta D_{2}^{ECH} = \Omega_{2} (d_{1}, c_{1})
\label{8}
\eeq
\beq
\delta D_{3}^{ECH} = \Omega_{3}(d_{1}, d_{2}, c_{1}, c_{2})
\label{9}
\eeq
\beq
\delta D_{4}^{ECH}=\Omega_4(d_1,d_2,d_3,c_1,c_2,c_3)
\label{new4}
\eeq
where $\Omega_{2}$, $\Omega_{3}$ and $\Omega_{4}$
are defined in Eqs. (11), (12) and (13) respectively.

One can understand from  Eqs. (6), (8) that the expressions
for $\Omega_N$ and
for the corrections $\delta D_N^{ECH}$ in Eqs. (\ref{8})-(17)
are the \underline{exact} numbers which are related to the special
scheme. This scheme is identical to the initial scheme at the
lower order levels and is defined by the condition $c_N=c_N^{ECH}$
at the ($N$+1)-order, where $c_N^{ECH}$ is considered as an
unknown
number. This means that the correction coefficients $\delta D_N$
are related to the initial scheme only partly. However, it was
shown in Refs. \cite{KatSt1,KatSt2}
 that in  certain cases the
numerical values of these coefficients are in  satisfactory
agreement with the results of the explicit calculations.
{\it A posteriori} we consider this fact as an argument in favour
of the possibility of the application of the re-expansion procedure
in the cases discussed by us.

In order to find similar corrections to Eq. (3) in the N-th order of
perturbation theory starting from the PMS approach \cite{PMS}, it is
necessary to use the relations obtained in Ref. \cite{yy} between the
coefficients $d_{i}^{PMS}$ and $c_{i}^{PMS} \; (i \geq 1)$ in the
expression for the order $O(a^{N}_{PMS})$ approximation
$D^{PMS}_{N} (a_{PMS})$ of the physical quantity under consideration:
\begin{equation}
d_i^{PMS}=\frac{1}{i+1}\bigg(\frac{N-2i-1}{N-1}\bigg)c_i^{PMS}
+O(a_{PMS})
\label{dipms}
\end{equation}
where $c_1^{PMS}=c_1$. Using now Eq. (\ref{dipms}) one can
find the corresponding coefficients of the NLO approximation
$D_2^{PMS}(a_{PMS})$
\begin{equation}
d_1^{PMS}=-\frac{1}{2}c_1+O(a_{PMS})
\label{d1pms}
\end{equation}
and the related expression for the NNLO correction
$\delta D_2^{PMS}$
\begin{equation}
\delta D_2^{PMS}=\delta D_2^{ECH}+\frac{d_0c_1^2}{4}
\label{deld2}
\end{equation}
where $\delta D_2^{ECH}$ is defined by Eqs. (15), (11).

Repeating now the similar considerations at   the NNLO
level we get from Eq. (\ref{dipms}) the following expressions
for the NLO and NNLO coefficients $d_1^{PMS}$ and $d_2^{PMS}$
\begin{eqnarray}
d_1^{PMS}&=& 0 +O(a_{PMS}); \nonumber \\
d_2^{PMS}&=& -\frac{1}{3}c_2^{PMS}+O(a_{PMS}).
\label{cornnlo}
\end{eqnarray}
 Substituting now Eqs. (\ref{cornnlo})
into Eq. (12) one can observe that the corresponding
next-to-next-to-next-to-leading order (N$^3$LO) correction
$\delta D_3^{PMS}$ in the re-expansion formula of Eq. (3)
identically coincide with $\delta D_3^{ECH}$ defined by
Eqs. (16), (12), namely that
\begin{equation}
\delta D_3^{PMS}=\delta D_3^{ECH} .
\label{d3cor}
\end{equation}
A similar observation
was made in Ref. \cite{KS}
 using different (but related) considerations.
In fact this expression means that the order $O(a_{PMS})$
correction to $d_1^{PMS}$ is cancelling the leading order
term in the expression for $d_2^{PMS}$. We have checked this feature
explicitely.

In  fourth order of the perturbation theory the additional
contribution to $\delta D_{4}^{PMS}$ has a more complicated structure.
In order to get it, it is necessary to substitute
 the following     expressions
\begin{eqnarray}
d_1^{PMS} &=& \frac{1}{6}c_1+O(a) \nonumber \\
d_2^{PMS} &=& -\frac{1}{9}c_2^{PMS}+O(a) \nonumber \\
d_3^{PMS} &=& -\frac{1}{4}c_3^{PMS}+O(a)
\label{pmsfour}
\end{eqnarray}
into the expressions for the scheme-invariants $\rho_2$
and $\rho_3$ and then into the analytical expression for
$\Omega_4$.
 The expression for $\Omega_4(d_i^{PMS},c_i^{PMS})$ in
Eq. (4), which results from this analysis, reads:
\beq
\Omega_4(d_i^{PMS},c_i^{PMS})=\frac{d_0}{3} [ \frac{1}{4}c_1c_3^{PMS}
-\frac{4}{81}(c_2^{PMS})^2-\frac{5}{81}c_1^2c_2^{PMS}+\frac{7}{648}
c_1^4]
\label{omega4}
\eeq
where
\begin{eqnarray}
c_2^{PMS}&=& \frac{9}{8}(c_2^{ECH}+\frac{7}{36}c_1^2) \nonumber \\
&=&\frac{9}{8}(d_2+c_2-d_1^2-c_1d_1+\frac{7}{36}c_1^2)
+O(a)
\label{c2pms}
\end{eqnarray}
and
\beq
c_3^{PMS}=4(d_3+\frac{1}{2}c_3-c_2d_1-3d_1d_2+2d_1^3)
+\frac{1}{2}c_1(d_2+c_2+3d_1^2-c_1d_1+\frac{1}{108}c_1^2)+O(a)
\label{c3pms}.
\eeq
The expressions for Eqs. (\ref{omega4})
- (\ref{c3pms}) are the pure numbers, which
do not depend on the choice of the initial scheme.
We will show in the next Section
that in the case of the consideration
of  perturbative series for  $(g-2)_{\mu}$
the numerical values of $\Omega_4(d_i^{PMS},c_i^{PMS})$ are
small and thus the {\it a posteriori} approximate equivalence
of the ECH and PMS approaches is preserved for the quantities
under consideration at this level also.

In  certain considerations we will need to use a generalization
of the expression for $\Omega_2$ to the case when the intitial
perturbative series is starting from  corrections of
order $O(a^p)$ with $p>1$
\begin{equation}
D_N^{(p)}=d_0a^p(1+\sum_{i\geq 1}\, d_i a^{N}).
\label{dp}
\end{equation}
In this case the expression for the corrections terms read
\begin{equation}
(\Omega_2^{(p)})_{ECH}
=\frac{p+1}{2p}d_0d_1^2+d_0d_1c_1 .
\label{omegapech}
\end{equation}
The  corresponding correction related to the
 PMS-improved expression was originally obtained
in Ref. \cite{PMS}:
\begin{equation}
(\Omega_2^{(p)})_{PMS} = \frac{p+1}{2p}d_0d_1^2+d_0d_1c_1+\frac{p}
{2(p+1)}d_0c_1^2.
\label{omegap}
\end{equation}

{\bf III.  Applications to $(g-2)_{\mu}$}

It is well-known that the expressions for  anomalous magnetic
moments of the electron $a_e=(g-2)_e/2$ and muon $a_{\mu}=(g-2)_{\mu}/2$
are known at the four-loop order from the results of calculations
of Ref. \cite{e} and Refs. \cite{mu1}, \cite{mu2} respectively.
The three-loop correction to $a_e$ is now  known with more
accuracy than previously \cite{Lap}. Combining the currently
available information about the coefficients of the perturbative
series for $a_e$ and $a_{\mu}$ we have the following expressions:
\begin{equation}
a_e= 0.5 a-0.3294789...a^2+1.17619(21)a^3-1.434(138)a^4
\label{ge}
\end{equation}
\begin{equation}
a_{\mu}-a_e=1.09433583(7)a^2+22.869265(4)a^3+127.55(41)a^4
\label{amu}
\end{equation}
where the expansion parameter $a=\alpha/\pi$ is related to the
fine structure constant $\alpha$ and the last term in Eq. (\ref{amu})
is the result of the
most recent calculations of Ref. \cite{mu2}
 stimulated by the work of Ref. \cite{BrKT}.
 Combining
Eq. (\ref{ge}) with Eq. (\ref{amu}) we arrive at the following
approximate expression for $a_{\mu}$:
\begin{equation}
a_{\mu}=0.5a+0.76585a^2+24 a^3+126a^4+O(a^5).
\label{amun}
\end{equation}
The order $O(a^5)$ correction to $a_{\mu}$ is only partly known
\cite{mu1}.
Our aim will be to try to touch the existing  uncertainty due to the
totally non-calculated order $O(a^5)$-contribution to Eq.
(\ref{amun}) using the
re-expansion procedure
outlined in the previous section
, which is compatible with the
RG-formalism.

 It is known that in the OS-scheme the
coefficients of the corresponding perturbative series depend on
the large $\ln(m_{\mu}/m_e)$-contributions starting from the two-loop
level. The parts of these effects are governed by the RG-method
\cite{BarRem,LR} (for a recent application of the RG method
to $a_{\mu}$, see Refs. \cite{FKLS,Kat3}). However, there are
also  certain $\ln(m_{\mu}/m_e)$-contributions, which are not
governed by the RG-method. They are associated with the
light-by-light-scattering electron loop
insertions coupled to the external
photon line. These contributions appear
first in the three-loop graphs,
which were subsequently calculated numerically in the works of
Ref. \cite{lblnum,mu1} and recently evaluated analytically in the work
of Ref. \cite{LapRem}.

In view of the different origin of the lower
$\ln(m_{\mu}/m_e)$-contributions we divide all diagrams into two
classes. The first class contains all diagrams with an
external muon vertex
and dressed interenal photon lines (see Fig. 1). As well as
in Ref. \cite{mu1} we will not  include the
diagrams with electron  loops to which four internal photon
lines
are attached. However, we will include four-loop diagrams typical
to $a_e$ but with substitution of the external electron vertex
to the muon one.
The second class
of the diagrams includes the diagrams with electron light-by-light
scattering
subgraph, to which three and four
internal photon lines are attached (see Fig.2).
Let us stress  that all $\log(m_{\mu}/m_e)$-terms of the
diagrams contributing to the first  class are totally controlled
by the RG-method, while in  class (II) only the parts of these
contributions
are governed
by the RG-technique.

In accordance with our classification we represent the
expression for $a_{\mu}$ in the following form
\begin{equation}
a_{\mu}=a_{\mu}^{(I)}+a_{\mu}^{(II)}.
\label{am}
\end{equation}
The concrete contributions to Eq.(33) read
\begin{eqnarray}
a_{\mu}^{(I)}&=&d_0^{(I)}a(1+d_1^{(I)}a+d_2^{(I)}a^2+
d_3^{(I)}a^3+...) \\
a_{\mu}^{(II)}&=&d_0^{(II)}a^3(1+d_1^{(II)}a+d_2^{(II)}a^2+...).
\end{eqnarray}
 Note, that the coefficients $d_i$ ($i\geq 1$) contain
the RG-controllable $\ln (x)=\ln(m_{\mu}/m_e)$-terms. Indeed, the
corresponding contributions to $a_{\mu}$ are governed by the
RG-equation
\begin{equation}
(m^2\frac{\partial}{\partial m^2}+\beta(a)\frac{\partial}{\partial a})
a_{\mu}^{(I,II)}=0
\label{rg}
\end{equation}
where $\beta(a)$ is the QED $\beta$-function in the OS-scheme, which
is defined as
\begin{equation}
m^2\frac{\partial a}{\partial m^2}=\beta(a)=\sum_{i\geq 0} \,
\beta_i a^{i+2}.
\label{beta}
\end{equation}
To our point of view, the separation of all diagrams to the two
classes mentioned above is respected by the property of the
RG-invariance. At least we do not know the arguments why the
 sum of the diagrams which belong to the class (I) and to
the class (II) should not obey the RG-equations seperately.

The coefficients of the $\beta$-function
  are known at the
four-loop level \cite{BrKT}. They have the following form
\newpage
\begin{eqnarray}
\beta_0&=&\frac{1}{3} \nonumber \\
\beta_1&=&\frac{1}{4} \nonumber \\
\beta_2&=&-\frac{121}{288}=-0.42 \nonumber \\
\beta_3&=&\bigg(\frac{5561}{5184}-\frac{23}{9}\zeta(2)
+\frac{8}{3}\zeta(2)\ln(2)-\frac{7}{8}\zeta(3)\bigg)\frac{1}{2}
\nonumber \\
&=&-0.571.
\label{coeff}
\end{eqnarray}
Thus, the related coefficients $c_i=\beta_i/\beta_0$ ($i\geq1$)
read $c_1=3/4$, $c_2=-1.26$, $c_3=-1.713$.
Let us write down the asymptotic expansions of the coefficients
of the contributions $a_{\mu}$ as
\begin{eqnarray}
d_0^{(I)}&=&B_1 \nonumber \\
d_0^{(I)}d_1^{(I)}&=& B_2+C_2 \ln(x) \nonumber \\
d_0^{(I)}d_2^{(I)}&=& B_3+C_3 \ln(x)+D_3 \ln^2(x) \nonumber \\
d_0^{(I)}d_3^{(I)}&=& B_4+C_4 \ln(x) +D_4 \ln^2(x) + E_4 \ln^3(x)
\nonumber \\
d_0^{(I)}d_4^{(I)}&=& B_5+C_5 \ln(x) +D_5 \ln^2(x)+E_5 \ln^3(x)
+F_5 \ln^4(x)
\label{dI}
\end{eqnarray}
and
\begin{eqnarray}
d_0^{(II)}&=&\overline{B}_1 \nonumber \\
d_0^{(II)}d_1^{(II)}&=&\overline{B}_2+\overline{C}_2 \ln(x) \nonumber \\
d_0^{(II)}d_2^{(II)}&=&\overline
{B}_3+\overline{C}_3 \ln(x)+\overline{D}_3 \ln^2(x) .
\label{dII}
\end{eqnarray}
The coefficients $C_i$, $D_i$, $E_i$, $F_i$ and $\overline{C}_i$,
$\overline{D}_i$ can be related to the coefficients of the
$\beta$-function using either the RG-considerations of Refs.
\cite{BarRem,LR,Kat3} or the explicit expressions for the
coefficients $\Omega_i$ and $\Omega_i^{(p)}$ in the corresponding
re-expansion formulas (see Eqs.(11)-(14) and Eq.(28)). The results
of the corresponding analysis have the following form
\begin{equation}
C_2=2\beta_0B_1
\end{equation}
\begin{eqnarray}
C_3&=&4\beta_0B_2+2\beta_1B_1  \nonumber \\
D_3&=&4\beta_0^2B_1
\end{eqnarray}
\begin{eqnarray}
C_4&=&6\beta_0B_3+4\beta_1B_2+2\beta_2B_1 \nonumber \\
D_4&=&12\beta_0^2B_2+10\beta_0\beta_1B_1  \nonumber \\
E_4&=&8\beta_0^3B_1
\end{eqnarray}
\begin{eqnarray}
C_5&=&8\beta_0B_4+6\beta_1B_3+4\beta_2B_2+2\beta_3B_1 \nonumber \\
D_5&=&24\beta_0^2B_3+28\beta_0\beta_1B_2+6\beta_1^2B_1
+12\beta_0\beta_2B_1 \nonumber \\
E_5&=&32\beta_0^3B_2+\frac{104}{3}\beta_0^2\beta_1B_1 \nonumber \\
F_5&=&16\beta_0^4B_1
\end{eqnarray}
\begin{eqnarray}
\overline{C}_2&=&6\beta_0\overline{B}_1 \nonumber \\
\overline{C}_3&=&8\beta_0\overline{B}_2+
6\beta_1\overline{B}_1 \nonumber \\
\overline{D}_3&=&24\beta_0^2\overline{B}_1 .
\end{eqnarray}
Note, that in the case of the diagrams of set (II)
 the corresponding coefficients
$\overline{B}_1$,
$\overline{B}_2$ and
$\overline{B}_3$  contain
the contributions of the non-controllable by the RG method
$\ln(x)$-terms.

Let us first discuss the applications of the procedure of Sec.II
to the diagrams of set (I). In this case the correction
terms $\Omega_2-\Omega_4$ reproduce all $\ln(x)$-contributions
presented in Eqs. (39). Moreover, one can get from re-expansion
procedure the exact values of the constant terms $B_i$ ($i\geq 3$)
 which do not depend on   the $\ln(x)$-terms.
In the case of the
application of the ECH-improved variant of the
OS-scheme these constant terms are defined by the conditions
\begin{eqnarray}
B_i&=&
\Omega_{i-1}(d_0^{OS},d_1^{OS},...,d_{i-2}^{OS},c_1,..,c_{i-2}^{OS})
\nonumber \\
&=&\Omega_{i-1}(B_1^{OS},...,B_{i-1},c_1,...,c_{i-2}^{OS}) .
\end{eqnarray}
 Similar terms which arise from the PMS-improved
expressions can be obtained after taking into account the
additional scheme-independent contributions derived in Sec.II.
We will demonstrate that the numerical values of these contributions
in the cases considered by us are not large.

The concrete values of the coefficients $B_1,B_2^{OS},B_3^{OS}$
  are known from a comparison of the results of the RG-inspired
analysis with the results of the
analytical and numerical calculations \cite{mu1}.
The coefficient $B_1=0.5$ is of course well known.
The asymptotic expression of the coefficient $B_2$, derived
in the limit $m_e/m_{\mu}\rightarrow 0$, can be found in Ref.
\cite{mu1}:
$B_2^{OS}=-\frac{25}{36}+a_e^{(4)}=-1.022923$.
The value of the coefficient $B_3^{OS}=2.741$ was obtained
in Ref. \cite{mu1} after subtracting
the contributions
of the light-by-light scattering graphs of the set (II) and of the
 RG-controllable contribution of Eq. (39)
from the expression
for the three-loop correction to $a_{\mu}$.

The value of the coefficient $B_4^{OS}$, which will be used by us,
is different from the one given in Ref. \cite{mu1}. The difference
comes from the fact that  contrary to the classification of
Ref. \cite{mu1},
we are including into the considered
set (I) the four-loop diagrams typical to $a_e$ but with
 substitution of the electron vertex and internal electron loops
to the muon ones.
 Moreover, it is necessary to modify the
 value of $B_4^{OS}$ presented in Ref. \cite{mu1}
in accordance with the results of
the analytical \cite{BrKT} and numerical \cite{mu2} re-calculations
of the diagrams with  three-loop insertion into the internal
photon line of the lowest order contribution to $a_{\mu}$.

In order to determine the value of the coefficient $B_4^{OS}$
in our case we used the following expression
\begin{equation}
B_4^{OS}=a_{\mu}^{(8)}-A_{\mu}^{(8)}(\gamma\gamma)
-C_4\ln(x)-D_4\ln^2(x)-E_4\ln^3(x)
\end{equation}
where $C_4$, $D_4$ and $E_4$ are determined by Eqs.(43) and
the value of $A_{\mu}^{(8)}(\gamma\gamma)\approx -116.7$
is the sum of the eight-order contributions of the diagrams
with electron light-by-light scattering subgraphs \cite{mu1}.
The numerical value of the coefficient $B_4^{OS}$ is thus
$B_4^{OS}=-7.74$.

In order to study the predictive abilities of the
 re-expansion procedure described in Sec. II we present in Table 1
the numerical results of our estimates of the coefficients
$B_i$ ($i \geq 3$) and
 compare them with  the  exact results for
$B_3^{OS}$ and $B_4^{OS}$ presented above.
\begin{center}
\begin{tabular}{|c|c|c|c|} \hline
Order & $B_i^{OS}$ & $B_i(ECH)$ & $B_i(PMS)$  \\ \hline
i=1 & 0.5 & ---- & ---  \\ \hline
i=2 & $-$ 1.022923 & --- & ---   \\ \hline
i=3 & 2.741 & 1.326 & 1.396  \\ \hline
i=4 & $-7.74$ & $-5.48$ & $-5.48$   \\ \hline
i=5 & --- & 41.6 & 41.7  \\ \hline
\end{tabular}
\end{center}
\vspace*{0.5mm}
Table 1:  Estimated values of the coefficients $B_i$ for
the diagrams of set (I).

One can see that the  re-expansion procedure used by us
is reproducing well enough the values of the coefficients
$B_3$ and $B_4$ (it gives the correct sign and predicts
the order of magnitude of these coefficients).
 Therefore, we hope that the estimate of the
five-loop constant term $B_5$ is also rather realistic.
Notice also the sign-alternating character of the results
of the estimates presented in Table 1. This feature has
something in common with the expectation that the
 RG-improved QED series for the Euclidean physical
quantities should have  sign-alternating behaviour \cite{asymp}.

Taking now into
account the numerical value        of the RG-controllable terms in
Eqs.(39), (44)  we arrive at the following estimate of the five-loop
contributions of the diagrams of set (I) into $a_{\mu}$
\begin{eqnarray}
a_{\mu}^{(10)} ( I) &=& B_5^{OS}+ 8.55\nonumber \\
&=& 50.1(ECH) \nonumber \\
&=& 50.2  (PMS)        .
\label{fini}
\end{eqnarray}
 This estimate is almost non-sensitive to the concrete
realization of the method of optimization. Notice also
the effect of reduction of  the value of the RG-controllable
 five-loop contributions presented
 in Refs. \cite{Kat3,BrKT}. Let us stress again that
this fact is explained by  necessity of the modifications
of the  results  used in Refs. \cite{Kat3,BrKT}  for the
constant term $B_4^{OS}$ derived in
 Ref.\cite{mu1}. These modifications
come from two  ingredients. First, it is necessary to use
 the corrected expressions
 obtained in Refs. \cite{BrKT,mu2}
of  certain four-loop graphs contributing to $a_{\mu}$
and second to add to the
 values of $B_4^{OS}$
cited in Ref. \cite{mu1}
the constant terms due to the four-loop graphs typical of
$a_e$ but with a substitution of the  electron
vertex and internal
electon loops to the muon ones. As is known from the results
of Ref. \cite{Lap5} the addition to the considerations of
Ref. \cite{Kat3} of the diagrams with the internal muon loops leads
to  strong cancellations. The comparison of the RG-controllable
expressions of Eq.(48) with the similar one derived in
Refs. \cite{Kat3,BrKT}
indicates the same pattern.

Let us now discuss the applications of the outlined procedure
for the estimates of the five-loop contributions of the diagrams
with the light-by-light scattering subgraphs of Fig.(2).
The special feature of the application of the re-expansion procedure
to the diagrams of set (II) is that the corresponding
terms $\overline{B}_i$
 depend from the $\ln(x)$-terms
 non-controllable by the RG method.
 The  most precise value of the coefficient
$d_{0}^{(II)}=\overline{B}_1=20.94792...$
is known from the
results of the analytical calculations of Ref. \cite{LapRem}.
The numerical result for the sum of the
corresponding four-loop graphs
reads \cite{mu1}
\begin{equation}
d_0^{(II)}d_1^{(II)}= 116.7  .
\label{lbliii}
\end{equation}
Using now Eqs. (28), (29) we arrive at the following numerical
estimate of the sum of the corresponding five-loop graphs
\begin{eqnarray}
a_{\mu}^{(10)}(II)=\Omega_2^{(3)} &=& 520.8(ECH)   \nonumber \\
&=& 525.2   (PMS)
\label{amuIII}
\end{eqnarray}
which includes the contribution of both
 RG-controllable and RG-non-controllable $\ln(x)$-terms.

The estimates of Eqs.(50), (48) should be compared with the
one given in Ref. \cite{mu1}
\begin{equation}
a_{\mu}^{(10)}=570(140)
\label{kine}
\end{equation}
where the central value comes from the exact calculation of
the contributions of the diagrams of the set of the light-by-light-type
diagrams with two one-loop electron loops inserted into the
internal photon lines (see Fig.3 ) and the error
bar $\pm 140$ stands for the estimate of other contributions
(mainly  RG-controllable ones).
Our estimate of Eq. (50) is in  very good agreement
with the central value of the estimate of Eq. (\ref{kine}),
while the estimate of Eq.(48) lies within the range of
 the
careful estimate $\pm 140$ of other contributions.

Note, however, that our total estimate
 of the considered tenth-order terms
\begin{eqnarray}
a_{\mu}^{(10)} &=& 570.9 (ECH) \nonumber \\
&=& 575.4 (PMS) .
\label{total}
\end{eqnarray}
also includes the contribution of the tenth-order
diagrams depicted in Fig.4 and not included in the
estimate of Eq. (\ref{kine}). These diagrams are formed by
the insertion of the two-loop electron loop into the internal
photon line of the lower light-by-light-type diagram.
The  contribution  of this set
of  tenth-order diagrams was
 estimated  in Ref. \cite{Karsh}.
In order to understand the uncertainties of this estimate
better
it is useful to write down
 a RG-relation analogous to Eqs. (43)
 for this set  of  diagrams separately.
Notice, that this contribution should be proportional to
the two-loop coefficient of the $\beta$-function (which is
determined by the graphs inserted into the internal photon
line). Using this observation we arrive at the following relation
\begin{equation}
a_{\mu}^{(10)}(Fig.4)= \overline{B}_3(Fig.4)
+6\overline{B}_1\beta_1\ln(x) .
\label{est2}
\end{equation}
The main contribution to the estimate of Ref. \cite{Karsh}
comes from the $\ln(x)$-term. Indeed, it has the following
numerical value $6\overline{B}_1\beta_1\ln(x)=167.47$.
This expression should be compared with the estimate
$a_{\mu}^{(10)}(Fig.4)=176\pm35$ given in Ref. \cite{Karsh}.
One can see that this estimate is relevant to the RG-controllable
contribution only. However, from the re-expansion procedure
we can see that the contributions
non-controllable by the RG-methods
might be non-negligible (see Eg. (\ref{fini})) and might affect
the final numerical value of the diagrams belonging to this set.
 In order to study this
guess in detail it is of interest to calculate the diagrams
of Fig.4 explicitly. This calculational project is rather
realistic \cite{Kinp}.

It is also interesting to understand deeper the uncertainties
due to
other diagrams which are  included neither in the
``optimized'' estimates of Eqs. (48), (50) nor in the original
estimates of Ref. \cite{mu1}. These diagrams, depicted
in Fig. 5, form a new class of diagrams, which cannot be
touched by the RG-inspired analysis. Indeed, one can hardly expect
that any resummation
procedures dealing with  light-by-light-type
graphs with three internal photon lines will be
able to give the estimate of the light-by-light-type
graph with five internal photon lines.
The expressions for the $\ln(x)$-terms
the non-controllable by the RG-method  for this
type of graphs can be
read from the considerations of Refs. \cite{newclass}. The
result was used in Ref. \cite{Karsh} where the following
estimate of the diagrams of Fig.5 was presented
\begin{equation}
a_{\mu}^{(10)}(Fig.4)=185\pm 85.
\label{add}
\end{equation}
Combining our estimates of Eq.(\ref{total})
with the ones of Eq.(\ref{add})
we get the final result of applications of the
re-expansion procedure
supplemented by the estimates of the
diagrams of new structure which are non-touched by this method
\begin{equation}
a_{\mu}^{(10)}\approx 700.
\label{final}
\end{equation}

Let us stress again that the new ingredient of our analysis, which
distinguishes it from  previous applications of
 the re-expansion
procedures in QED
\cite{PMS,KS,F}, is the  separation of the considered initial diagrams
to two classes, one of which consists of the diagrams, relevant
to the effects of ``new physics", discussed in more detail in
Refs. \cite{Yelkh,newclass}. This procedure finds its support
in the theoretical considerations of Ref.\cite{BLM}.

Moreover, we checked
 that in spite of the good agreement
of the application of the re-expansion procedure to the non-separated
sixth-order expressions for $a_{\mu}$ with  results of the
eighth-order calculations \cite{KS}, the straightforward
application of Eq.(13)
 to the non-separated
eighth-order approximation of Eq. (32) results in the non-confortably
large tenth-order estimate $a_{\mu}^{(10)}\approx 2160$.
It is possible to understand that the reason of the success of the
application of
the re-expansion procedure to the non-separated sixth-order approximation
is connected with the fact that the use  of Eq.(12) (and more
definitely its last term)  gives
 for the eight-order light-by-light-type term the following estimate
$a_{\mu}^{(8)}(\gamma\gamma)=3d_0d_1d_2(\gamma\gamma)=
6a_{\mu}^{(4)}a_{\mu}^{(6)}(\gamma\gamma)$
which is known to be in good agreement with the results of direct
numerical calculations \cite{mu1}. However, at the next level
of perturbation theory the expression for the correction term
$\Omega_4$ of Eq.(13) has a more complicated structure and thus
the resulting non-separated estimates turn out to be
non-confortably large. Moreover, we consider the satisfactory
agreement of the results of separated estimates with the estimates
given in Ref. \cite{mu1} as an argument in favour of treating
the diagrams with the electron-loop
light-by-light scattering graphs separately.

Another interesting question is connected with the problem
of the comparison of our estimates with the results of the
recent applications of the Pad\'e resummation technique
to the perturbative series for $a_{\mu}$ \cite{SLS} and
$a_{\mu}-a_e$ \cite{SLS,EKSS}. It should be stressed that
in their analysis
the authors of Refs. \cite{SLS,EKSS} did not consider
the light-by-light scattering graphs separately. Note
also that the
 coefficients of the corresponding Pad\'e approximants
depend from the $\ln(x)$-terms. In spite of the fact
that our results for $a_{\mu}$ are in qualitative agreement
with the results of the applications of the Pad\'e resummation
method  \cite{SLS,EKSS}
it is   interesting
to try to understand the predictive abilities of the Pad\'e
resummation methods better. Clearly, this problem is connected
with the necessity of more detailed understanding of the
relations of the Pad\'e results to the ones obtained using
the RG-inspired analysis. Note, that the
  Pad\'e resummation methods can face the problem in reproducing
the structure of the RG-controlable $\ln(x)$-terms.
{\bf Acknowledgements}

It is a pleasure to thank
 R.N. Faustov, T. Kinoshita and P.M. Stevenson for discussions.

{\bf Appendix}

Using Eq.(14) it is possible to derive the six-loop RG-controllable
contributions to the diagrams of set (I):
\begin{equation}
d_0^{(I)}d_5^{(I)}=B_6+C_6\ln(x)+D_6\ln^2(x)+E_6\ln^3(x)
+F_6\ln^4(x)+G_6\ln^5(x) .
\end{equation}
The expressions for the logarithmic coefficients are
\begin{eqnarray}
C_6&=&10\beta_0B_5+8\beta_1B_4+6\beta_2B_3+4\beta_3B_2+2\beta_4B_1
\nonumber \\
D_6&=&40\beta_0^2B_4+54\beta_0\beta_1B_3+16\beta_1^2B_2
+32\beta_0\beta_2B_2+14\beta_1\beta_2B_1+14\beta_0\beta_3B_1
\nonumber \\
E_6&=& 80\beta_0^3B_3+\frac{376}{3}\beta_0^2\beta_1B_2+
\frac{140}{3}\beta_0\beta_1^2B_1+48\beta_0^2\beta_2B_1 \nonumber \\
F_6&=& 80\beta_0^4B_2+\frac{308}{3}\beta_0^3\beta_1B_1 \nonumber \\
G_6&=& 32\beta_0^5B_1.
\end{eqnarray}
However, in order to use these expressions in  concrete considerations
it is necessary to fix somehow the value of the five-loop coefficient
$\beta_4$
of the QED $\beta$-function in the OS scheme.

\end{document}